\documentclass[11pt]{article}
\usepackage{array}
\usepackage{amsmath}
\usepackage{amssymb}	
\newcommand{\vev}[1]{\left\langle #1 \right\rangle}

\begin{document}
\begin{titlepage}
\begin{center}
\hfill arXiv:\,0802.0009\\ 
\vskip 1cm
\begin{LARGE}
\textbf{Supergravity on $\mathbb{R}^{4}\times S^{1}/\mathbb{Z}_{2}$ and singular Calabi-Yaus}
\end{LARGE}\\
\vspace{1.0cm}
Sean McReynolds \footnote{sean.mcreynolds@mib.infn.it}\\
\vspace{.35cm}
\emph{University of Milano-Bicocca and INFN Milano-Bicocca\\
Piazza della Scienza 3, 20126 Milano, Italy}\\
\vspace{1.0cm} {\bf Abstract}
\end{center}
We discuss the moduli space singularities that are generally present in five-dimensional vector-coupled supergravity on a spactime of the form $\mathbb{R}^{4}\times S^{1}/\mathbb{Z}_{2}$, with vector fields surviving on the $\mathbb{Z}_{2}$ fixed planes.  The framework of supergravity is necessarily ambiguous when it comes to the non-singular embedding theory, so we focus on those models coming from Calabi-Yau three-folds with wrapped membranes.  

\vfill {\flushleft {January 2008}}
\end{titlepage}

\section{Introduction}\label{sec:Introduction}
Supergravity theories in D-dimensions can have moduli space singularities, which may have a higher dimensional interpretation via 11D supergravity compactified on singular (11-D)-manifolds. Previous work has studied the framework of 5D supergravity theories resulting from singular Calabi-Yaus~\cite{MZ, effectiveflops}, with prior and later applications in cosmology~\cite{Gaida:1998km} and 5D heterotic M-theory~\cite{Kallosh:2000rn}, the latter involving a singular Calabi-Yau at a bulk point of the $S^{1}/\mathbb{Z}_{2}$ interval. We are interested in the fact that 5D Maxwell-Einstein supergravity on a spacetime of the form $M_{4}\times S^{1}/\mathbb{Z}_{2}$ can have such singularities on the $\mathbb{Z}_{2}$ fixed planes. As described in~\cite{MZ}, this requires an embedding in a new theory in which additional supermultiplets are massless on the planes. For the purpose of a self-contained exposition, in this section we review 5D supergravity, Calabi-Yau origins and the types of singularities that we are interested in.  We then consider the situation on spacetimes of the form $\mathbb{R}^{4}\times S^{1}/\mathbb{Z}_{2}$.     

Minimal ($\mathcal{N}=2$) 5D supergravity, consisting the ``bare"  multiplet
\[
\{g_{\hat{\mu}\hat{\nu}},\Psi^{i}_{\hat{\mu}},A^{0}_{\hat{\mu}}\},
\]       
can be coupled to $n_{V}$ ``bare" abelian vector supermultiplets
\[
\{A^{i}_{\hat{\mu}},\lambda^{p\,\ell},\phi^{x}\},
\] 
(where $\hat{\mu}=0,1,2,3,5$; $i=1,\ldots,n_{V}$; $x=1,\ldots, n_{V}$; and $\ell=1,2$ is an $SU(2)_{R}$ index)
to form Maxwell-Einstein supergravity theory (MESGT), with bosonic Lagrangian~\cite{GST}
\begin{equation}\begin{split}
e^{-1}\mathcal{L}_{5}=&-\frac{1}{2\kappa^{2}}\mathcal{R}-\frac{1}{4}\stackrel{\circ}{a}_{IJ}F^{I}_{\hat{\mu}\hat{\nu}}F^{J\;\hat{\mu}\hat{\nu}} - \frac{3}{4\kappa^{2}}\stackrel{\circ}{a}_{IJ}\partial_{\hat{\mu}}h^{I}\partial^{\hat{\mu}}h^{J}\\
&+\frac{\kappa e^{-1}}{6\sqrt{6}}C_{IJK}\epsilon^{\hat{\mu}\hat{\nu}\hat{\rho}\hat{\sigma}\hat{\lambda}}F^{I}_{\hat{\mu}\hat{\nu}}F^{J}_{\hat{\rho}\hat{\sigma}}A^{K}_{\hat{\lambda}}+\cdots
\end{split}\label{lagrangian}\end{equation} 
where $h^{I}$ and $\stackrel{\circ}{a}_{IJ}$ are functions of the scalars $\phi^{x}$ described below, $C_{IJK}$ is a constant symmetric tensor that completely defines the MESGT and the ellipsis indicates fermionic couplings.

Following~\cite{GST}, the target space geometry can be described as follows.  Consider the set $\Omega\subset\mathbb{R}^{n_{V}+1}$ consisting those $\xi^{I}$ satisfying $\mathcal{V}:=C_{IJK}\xi^{I}\xi^{J}\xi^{K}=e^{k}>0$, where $k\in \mathbb{R}$ parametrizes a foliation of hypersurfaces. The real $n_{V}$-dimensional scalar manifold $\mathcal{M}_{V}$ is the hypersurface defined by $k=0$; the $h^{I}(\phi)$ are proportional to the embedding functions $\xi^{I}|_{\mathcal{V}=1}$. The dependence of the $(n_{V}+1)$ functions $h^{I}$ on the $n_{V}$ scalars $\phi^{x}$ can be derived from the hypersurface condition.  The transformations leaving $C_{IJK}$ invariant form a (possibly trivial) rigid symmetry group $G$ of the Lagrangian,\footnote{There is also a rigid $SU(2)_{R}$ symmetry of the superalgebra that we ignore.} which is a subgroup of the isometry group $Iso(\mathcal{M}_{V})$ of the scalar manifold.  On $\Omega$, a Riemannian metric can be defined as $a_{IJ}=-\partial^{2}\ln\mathcal{V}/\partial\xi^{I}\partial\xi^{J}$, while the metric appearing in the kinetic terms of Eq.~(\ref{lagrangian}) is the restriction to the scalar manifold: $\stackrel{\circ}{a}_{IJ}=a_{IJ}|_{\mathcal{V}=1}$. The set of $\stackrel{\circ}{a}_{IJ}$ belong to a cone $\mathcal{K}$ of non-degenerate positive metrics; the boundary of $\mathcal{K}$ is where these metrics degenerate (or become singular). 

There are bases, which we call adapted, in which the cone $\mathcal{K}$ corresponds to $\xi^{I}>0$ for all $I$, while $\xi^{I}=0$ for any value of $I$ correspond to its boundary. Restricting to the part of the boundary such that $\mathcal{V}$ remains finite and positive definite, one can construct a supergravity theory that is well-defined.  The expected interpretation is that the singularities in the original theory's moduli space arise from degrees of freedom that become massless at those points.  

In order to understand the situation, it helps to turn to 11D supergravity compactified on Calabi-Yau spaces.  While 5D supergravity theories have not been proven in general to be low energy descriptions of string/M-theory, specific theories have been shown to admit such a description.  If a 5D MESGT can be found from 11D sugra on a smooth Calabi-Yau 3-space, $X$, the theories are related as follows~\cite{CYcompactification, AntoCY}:\footnote{There will also be hypermultiplets appearing in the smooth compactification, but we leave details to the Appendix.}  
\begin{itemize}
\item
The polynomial $\mathcal{V}(h)=1$ is a rescaling of $X$'s volume $\tilde{\mathcal{V}}=\int_{X} J\wedge J\wedge J$ by itself, where $J$ is the K\"{a}hler 2-form of $X$.  
\item
In the adapted basis, the functions $h^{I}= \tilde{h}^{I}/\tilde{\mathcal{V}}^{1/3}$ are rescalings of holomorphic 2-cycle volumes 
\[\tilde{h}^{I}=\int_{\mathcal{C}^{I}} J=vol(\mathcal{C}^{I})>0,\]                       
where the $\mathcal{C}^{I}$ form a basis of $H_{2}(X;\mathbb{Z})$.  
\item
The $h_{I}=\tilde{h}_{I}/\tilde{\mathcal{V}}^{2/3}$ are rescalings of holomorphic 4-cycle volumes
\[
\tilde{h}_{I}=\int_{\mathcal{D}_{I}}J\wedge J=vol(\mathcal{D}_{I})>0,
\]
where the $\mathcal{D}_{I}$ form a basis for $H_{4}(X;\mathbb{Z})$.
\item
The $C_{IJK}$ are triple intersection numbers 
\[
C_{IJK}= \mathcal{D}_{I}\circ \mathcal{D}_{J}\circ \mathcal{D}_{K}. 
\]
\end{itemize}
Note that the $C_{IJK}$ are integer valued in the basis in which we've defined them.  The 2- and 4-cycles are dual in that their intersections satisfy $\mathcal{C}^{I}\circ\mathcal{D}_{J}=\delta^{I}_{J}$.

In the class of theories arising as compactifications on a Calabi-Yau, then, the boundary of the cone $\mathcal{K}$ of target space metrics corresponds to the (classical) boundary of the cone of Calabi-Yau K\"{a}hler metrics; in an adapted basis, $h^{\star}\rightarrow 0$ (where $\star$ is some value of $I$) corresponds to the collapse of a 2-cycle $\mathcal{C}^{\star}$. Naturally, one can consider the BPS-extended superalgebra of 11D sugra including charges for 2- and 5-branes.  These objects can wrap 2- and 4-cycles (respectively) of a Calabi-Yau, yielding electric 0-branes and magnetic 1-branes associated with charges in the BPS-extension of the 5D superalgebra~\cite{AntoCY, 5dBPS}.  The mass of these 5D objects depends on the volumes of the cycles they wrap; in particular, a membrane wrapping a \textit{holomorphic} 2-cycle $\mathcal{C}^{\star}$ has vanishing mass as $h^{\star}\rightarrow 0$.\footnote{We are assuming that the CY volume $\tilde{\mathcal{V}}$ remains finite and non-zero in the process.}  From the 5D point of view, these point-like BPS objects provide the new charged field content that we should include~\cite{Strominger:1995cz}.

The nature of the new field content depends on the nature of the singularities.  At the boundary of a cone in which a 2-cycle collapses, one can move into a new cone by blowing up a 2-cycle or a 3-cycle~\cite{flops}. We'll consider several types of transitions in which the singularity is blown up with a 2-cycle~\cite{KMP, KM, Witten:1996qb}.  First, if the homology class of the contracted cycle $\mathcal{C}^{\star}$ contains a \textit{finite number} $\aleph$ of isolated holomorphic curves, the transition is a flop; flopped spaces are related by a mapping $\rho: \mathcal{C}^{\star}\rightarrow -\mathcal{C}^{\star}$~\cite{flops}. At the singularity, there will be $\aleph$ new hypermultiplets charged with respect to the $U(1)$ vector field $A^{\star}_{\hat{\mu}}$~\cite{KMP, KM, Witten:1996qb}.  

Alternatively, the contracted 2-cycle $\mathcal{C}^{\star}$ may consist a \textit{continuous family} of holomorphic curves, itself parametrized by a curve of genus $g$ to which the complex surface that the family sweeps out collapses. This family is part of a whole class of complex surfaces $\mathcal{D}\in H_{4}(X;\mathbb{Z})$, called an exceptional divisor, which is induced to collapse to a class $\mathcal{C}\in H_{2}(X;\mathbb{Z})$ of genus $g$ curves (which are called \textit{exceptional} singular curves, though we may drop this).  The associated divisor is a linear combination of basis 4-cycles $\mathcal{D}=v^{I}\mathcal{D}_{I}$, where $v^{I}\in \mathbb{Z}$, satisfying $\mathcal{D}\circ \mathcal{C}^{\star}=-2$. A divisorial collapse to a curve can be characterized by $\lim_{h^{\star}\rightarrow 0}v^{I}h_{I}\sim h^{*}$~\cite{MZ}. The transition in this case is called elementary and is characterized by the mapping $\epsilon:\mathcal{D}\rightarrow -\mathcal{D}$ (see~\cite{KMP,KM} and references therein).  The curve of singularities locally sees a transverse space of the form $\mathbb{C}^{2}/\mathbb{Z}_{2}$.  There may be points along the exceptional singular curve that are higher order, seeing a local transverse space  $\mathbb{C}^{3}/\Gamma_{n}$, where where $\Gamma_{n}$ is a rank-n discrete group with $n>2$.  These are called exceptional singular points, and tend to be blown up with complex surfaces that admit a special rational (genus 0) curve; contracting a particular 2-cycle over this point allows a special divisorial collapse that looks like the situation above.    
  
Upon divisorial contraction to a genus $g$ curve of $A_{1}$ singularities (locally $\mathbb{C}^{2}/\mathbb{Z}_{2}$),   the roots of $A_{1}$ correspond to new 5D $\mathcal{N}=2$ vector multiplets charged with respect to $A^{\star}_{\hat{\mu}}$; together they form an $SU(2)$ gauge multiplet.  There will also be $g$ 5D $\mathcal{N}=2$ hypermultiplets in the adjoint representation of the enhanced $A_{1}$ symmetry group~\cite{KMP,KM,Witten:1996qb}.  This is encoded in the $C_{IJK}$ tensor of the 5D theory since~\cite{KMP}  
\begin{equation}
\mathcal{D}^{3}\equiv v^{I}v^{J}v^{K}C_{IJK}=8(1-g)\;\;\;\;\;\;\mbox{CY Adapted Basis}.\label{dcubed}
\end{equation}
In other bases this formula changes by a proportionality factor. 

In general, a set of $h^{\alpha}\rightarrow 0$ can cause multiple divisors $\mathcal{D}_{(\alpha)}=v^{I}_{(\alpha)}D_{I}$ ($\alpha=1,\ldots, n$) to collapse to a singular curve of genus $g$. If the intersection matrix of the $\mathcal{D}_{(\alpha)}$ corresponds to a Dynkin diagram of a larger $A_{n}$, $D_{n}$ or $E_{n}$ type group, then there will be new charged vector multiplets corresponding to roots, and $g$ hypermultiplets forming the adjoint representation of that group~\cite{KMP,KM,Witten:1996qb}.  In addition, though we don't consider it here, the contraction of multiple divisors allows situations with a non-abelian generalization of the flop contraction in which only $n-1$ of the $n$ divisors are ruled surfaces (sweeping out a continuous family of 2-cycles), with the remaining one being a finite collection $\aleph$ of 2-cycles in the same homology class. Upon collapse of the divisors one gets an $A_{n}$, $D_{n}$ or $E_{n}$ gauge group again, but with $\aleph$ hypermultiplets in the \textit{fundamental} representation~\cite{NewHiggs,Intriligator:1997pq}. The generalization of Eq.~(\ref{dcubed}) to these situations is discussed in~\cite{Intriligator:1997pq}.

In any case, compactification of M-theory on singular Calabi-Yaus of these types yield 5D Yang-Mills-Einstein supergravity theories (YMESGTs) in the long-wavelength limit.  YMESGTs~\cite{YMESGTs} are obtained from MESGTs if one can gauge a subgroup $K\subset G$ of the rigid symmetry group.  This requires that a subset of the vectors fall into the adjoint representation of $K$. With remaining vectors falling into other representations, and it must be possible to dualize any charged vectors to tensors.  

We can now turn to the effective supergravity description of these transitions, following~\cite{MZ}. We wish to obtain a new theory based on a polynomial $\hat{\mathcal{V}}$ from the original $\mathcal{V}$.  ``Integrating in" additional hypermultiplets and vector multiplets charged with respect to an enhanced group changes the polynomial $\mathcal{V}$ by (keeping $h^{\alpha}>0$)~\cite{Intriligator:1997pq}  
\[
\delta \mathcal{V}_{in}=\frac{1}{2}(\sum_{f}\sum_{\mathbf{w}\in W_{f}}|\mathbf{w}\cdot\mathbf{h}|^{3}-\sum_{\mathbf{r}} |\mathbf{r}\cdot \mathbf{h}|^{3})
\]           
where $W_{f}$ is the set of weight vectors $\mathbf{w}$ for a given representation of the group labeled by $f$ (corresponding to new hypermultiplets); $\mathbf{r}$ are root vectors of the group (corresponding to the new vector multiplets); and $\mathbf{h}$ is the vector of scalar functions $h^{\alpha}$ that the new field content is charged with respect to.\footnote{In the case of a single contraction of $\mathcal{C}^{\star}$ yielding $N_{H}$ hyper- and $N_{V}$ vector multiplets, we simply have  $\delta\mathcal{V}_{in}=\frac{1}{2}(N_{H}-N_{V})(h^{*})^{3}$~\cite{AntoCY,  Witten:1996qb, integratedout}.}  In the case of divisorial collapse, the non-singular emedding theory should have additional vector multiplets, so there must also be new scalar functions $h^{A}$ ($A=1\ldots,N_{V}$) such that the new polynomial $\hat{\mathcal{V}}(h^{I},h^{A})$ admits a new A-D-E symmetry group.  For generic backgrounds of $h^{A}$, $\hat{\mathcal{V}}$ should describe a theory with massive BPS multiplets, while the background $h^{\alpha}=0=h^{A}$ allows enhanced A-D-E symmetry, in which case the multiplets have become un-Higgsed.        
The polynomials defining the vector-coupled sectors of the theories are then related by~\cite{MZ}
\[
\hat{\mathcal{V}}|_{h^{A}=0}=\mathcal{V}+\delta\mathcal{V}_{in},
\]
where the left hand side is simply a truncation. One must then extend this to the untruncated polynomial by adding terms involving the $h^{A}$ such that the enhanced symmetry of $\hat{\mathcal{V}}$ is preserved. 

Before we end this section, we make a technical clarification. In the adapted Calabi-Yau basis of the original theory, the condition $\mathcal{V}=1$ is violated for the subsurface $h^{\alpha}=0$, as well as for the corresponding points in other bases.  In obtaining the new theory, we imagine deforming the \textit{non-rescaled} Calabi-Yau volume $\tilde{\mathcal{V}}$ (see Section 1) to one for the ``generalized" Calabi-Yau $\hat{\tilde{\mathcal{V}}}(\tilde{h}^{I},\tilde{h}^{A})$; then we rescale using the latter to obtain the functions $h^{I}$, $h^{A}$ and the new polynomial condition $\hat{\mathcal{V}}=1$. Now the condition $h^{\alpha}=0=h^{A}$ corresponds to a phase of the ``generalized" Calabi-Yau that is not singular.  This will be true in other bases we choose, so that the new condition $\hat{\mathcal{V}}=1$ \textit{is} satisfied. We will remark on this in an example of Section~\ref{sec:Examples}. 
           
\section{MESGT on $S^{1}/\mathbb{Z}_{2}$}\label{sec:MESGT}

We now put 5D MESGT, with rigid symmetry group $G$, on $\mathbb{R}^{4}\times S^{1}/\mathbb{Z}_{2}$ in which the $\mathbb{Z}_{2}$ action on $S^{1}$ is represented by $x^{5}\rightarrow -x^{5}$, with $x^{5}\in [-\pi R, \pi R]$.  The fixed points are at $x^{5}=\{0\},\{\pm\pi R\}$. The $\mathbb{Z}_{2}$ action is then lifted to the $G$-bundle so that objects are also assigned parities according to their $G$-indices~\cite{parities}. In particular, some set of the 4D vector fields $A^{\alpha}_{\mu}$ ($\alpha:1,\ldots,n$; $\mu=0,\ldots,3$) are given even parity with respect to the fixed points so that they have propagating modes there.\footnote{The 5D graviphoton is always projected to a 4D scalar~\cite{parities}, but it does not play a direct role here.}  Supersymmetry then requires that the scalar functions, $h^{\alpha}(\phi)$, have odd parity. \textit{A priori} these functions may consist $C^{0}$ and $C^{-1}$ pieces. However, it is usually suggested that these functions go to zero (to be $C^{0}$) at the fixed points so as to avoid the appearance of $\delta(0)$ in the action (squared Dirac distributions in the Lagrangian and equations of motion).  For the class of theories we are discussing, arising from compactifications of 11D supergravity on a Calabi-Yau space, there is another motivation for the $C^{0}$ restriction: If we maintain that the $h^{\alpha}$ are proportional to volumes of a \textit{linearly independent} set of 2-cycles, they must contract. Technically, since the fixed planes are four-dimensional, we should be considering the complexified K\"{a}hler cone of Calabi-Yau spaces, in which the boundary is reached upon the vanishing of the \textit{complex} 4D scalar whose real and imaginary parts are  $A^{\alpha}_{5}$ and $h^{\alpha}$, respectively (the former are the ``theta angles")~\cite{KMP}. Although the $A^{\alpha}_{5}$ have odd parity, they are allowed to be $C^{-1}$ and so can jump across the fixed planes~\cite{McR2} with the $\theta(x^{5})$ distribution mentioned above.  In the isomorphic picture where the spacetime is a manifold with boundaries, this implies that it is possible for these fields to have non-vanishing boundary conditions.  However, the $\vev{A^{\alpha}_{5}}$ do vanish for supersymmetric backgrounds.  

Therefore, these 5D supergravity theories have a (now complex) codimension-$n$ surface of singularities in the moduli space, which is reached at the $\mathbb{Z}_{2}$ fixed points of the spacetime. Once we determine what new field content the theory has in five dimensions, we can assign $\mathbb{Z}_{2}$ parities to it in a consistent way to obtain the 4D $\mathcal{N}=1$ multiplets that are massless on the fixed planes.
   
For $\mathcal{V}$ to be parity-even, the components of $C_{IJK}$ with an odd number of $\alpha$ indices have odd parity, which we implement with the jumping function $\theta(x^{5})$, with value $-1$ for $(-\pi R,0)$ and $+1$ for $(0,\pi R)$.
This mimics the situation on either side of a flop and elementary singularity reviewed in the previous section: There the polynomials are of relatively different form because of the jumping coefficients, which result from the transitions $C^{\alpha}\rightarrow -C^{\alpha}$, but they describe equivalent theories since one also flips the signs of the $h^{\alpha}$. 
      
Following the discussion at the end of the previous section, the scalar background will be such that the new fields  have an $x^{5}$-dependent mass that is non-zero for points in the bulk spacetime, and vanishes on the fixed planes.     
For flop transitions, the 5D hypermultiplets become $\aleph$ chiral $\mathcal{N}=1$ multiplets on the fixed planes (charged with respect to formerly Maxwellian vector fields).  For elementary transitions, the expected symmetry enhancement on the fixed planes must be broken to a proper, rank-preserving subgroup; there is otherwise a choice in the assignment of $\mathbb{Z}_{2}$ parities, splitting 5D vector multiplets into either vector or chiral $\mathcal{N}=1$ multiplets on the fixed planes (the latter forming a real representation under the surviving gauge group).   Also on the fixed-planes, the 5D hypermultiplets form $g$ chiral $\mathcal{N}=1$ multiplets in the adjoint representation of the would-be full enhanced group.

\section{Examples}\label{sec:Examples}

I.  Consider the 2-moduli family of reducible polynomials arising from K3-fibered Calabi-Yaus~\cite{Hosono}
\[
\mathcal{V}=C_{000}(h^{0})^{3}+3C_{001}(h^{0})^{2}h^{1}
\]
where $(C_{000},C_{001})$ take the values in the table below. Their properties and singular transitions were discussed in~\cite{KM, KMP}.\\
\begin{center}\begin{tabular}{c|cccccc}\hline\\
$(C_{000},C_{001})$&(8,4)&(2,4)&(4,2)&(12,6)&(16,8)&(5,4)\\
$\mathcal{D}^{3}$&-16&-256&-8&-24&-32&--\\
$g$&3&65&2&4&5&--\\ \hline
\end{tabular}\end{center}
The dual $h_{I}$ are 
\[
h_{0}=C_{000}(h^{0})^{2}+2C_{001}h^{0}h^{1},\;\;\;\;
h_{1}=C_{001}(h^{0})^{2}.
\]
Let us first consider all but the last case in the table. These models undergo an elementary contraction since one can find a $\mathcal{D}=v^{I}\mathcal{D}_{I}$ such that $\mathcal{D}\circ C^{1}=-2$ (of course, any multiple or divisor of this $\mathcal{D}$, yielding integral $v^{I}$, vanishes).  Furthermore, each of these singularities differs in nature, as is reflected in the value of $\mathcal{D}^{3}\equiv v^{I}v^{J}v^{K}C_{IJK}$ in the table.  
Using Eq.~(\ref{dcubed}), this is consistent with the genera of the singular curves to which the single divisors collapse~\cite{KMP}. 
 We therefore know the additional field content in five dimensions: two new vectors join the vector $A^{1}_{\hat{\mu}}$ in the adjoint representation of $SU(2)$, along with $g$ hypermultiplets in the adjoint. 
 
 It's often convenient to use the canonical basis in which 
\[
C_{000}=1,\;\;\;\;C_{00i}=0,\;\;\;\;C_{0ij}=-\frac{1}{2}\delta_{ij},
\]
with the remaining $C_{ijk}$ specifying the type of theory.  
The vector-coupled sector of all the theories in the table are physically equivalent to a single model in the canonical basis:
\[
\mathcal{V}=(\check{h}^{0})^{3}-\frac{3}{2}\check{h}^{0}(\check{h}^{1})^{2}-\frac{1}{3\sqrt{6}}(\check{h}^{1})^{3}
\] 
via the basis transformation
\[
h^{0}=\frac{1}{C^{1/3}_{000}}\check{h}^{0}+\frac{1}{\sqrt{6}C_{000}^{1/3}}\check{h}^{1},\;\;\;\;\;\;
h^{1}=-\frac{C_{000}^{2/3}}{\sqrt{6}C_{001}}\check{h}^{1}.
\]
However, we see that the singularity structures of these theories as $h^{1}\rightarrow 0$ differ.  Note that, in the canonical basis, the condition $\mathcal{V}=1$ implies that there is a ``canonical basepoint"~\cite{GST} given by $(\check{h}^{0},\check{h}^{1})=(1,0)$ corresponding to ground states with maximal symmetry. However, as mentioned at the end of Section~\ref{sec:Introduction}, the truncation here does not correspond to moving to this point but generally off the $\mathcal{V}=1$ surface. 

Introducing two new (vector multiplet) scalar functions $h^{A}=\{h^{2},h^{3}\}$, the new polynomial has the simple form 
\[
\hat{\mathcal{V}}=(\check{h}^{0})^{3}-\frac{3}{2}\check{h}^{0}\sum_{i=1,2,3}\check{h}^{i}\check{h}^{i}.
\] 
Of course, this is the only form allowed by the $SU(2)$ symmetry of the enlarged theory since there are only symmetric rank-3 invariants for $SU(N)$, $N\geq 3$.
The simple $C_{ijk}=0$ class of theories correspond to non-homogeneous reducible scalar manifolds~\cite{GST}. Again, recalling the comments at the end of Section 1, the new polynomial is now considered to properly satisfy $\hat{\mathcal{V}}=1$ so that $h^{i}\rightarrow 0$ corresponds to the canonical basepoint $(\check{h}^{0},\check{h}^{i})=(1,0)$, which is a regular point of the theory. 

In these examples we start with a 5D MESGT with one Maxwell vector multiplet. On $\mathbb{R}^{4}\times S^{1}/\mathbb{Z}_{2}$, with $A^{1}_{\mu}$ given even $\mathbb{Z}_{2}$ parity, the theory becomes a $C_{ijk}=0$ type YMESGT, with three vector multiplets gauging the $SU(2)$ isotropy group of the corresponding scalar manifold, coupled to hypermultiplets.  The masses of the new multiplets (coming from backgrounds of $h^{2},h^{3}$) are $x^{5}$-dependent, vanishing on the fixed planes. The massless theory on the fixed planes consists a $U(1)$ YMESGT with $2g+2$ charged and $g$ neutral chiral multiplets (in addition to a neutral chiral multiplet from the 5D gravity multiplet that we have ignored throughout).

 The last entry in the table corresponds to a flop singularity.  In this case, it's known that there are $\aleph=16$ hypermultiplets charged with respect to $A^{1}_{\hat{\mu}}$, and no new vector multiplets~\cite{N16flop,KM}.  The new polynomial is now
\[
\hat{\mathcal{V}}=(\check{h}^{0})^{3}-\frac{3}{2}\check{h}^{0}(\check{h}^{1})^{2}+\frac{3}{16\sqrt{6}}(\check{h}^{1})^{3}.
\]        
The 4D theory on the boundaries will have 16 chiral multiplets charged with respect to $A^{1}_{\mu}$. There appears to be a freedom in whether we project a given hypermultiplet to a $+$ or $-$ charge chiral multiplet. But ultimately, one must understand the hypermultiplet sector in terms of a quaternionic target manifold~\cite{BaggerWitten, hypers} that admits specific isotropy groups.  The representations of the hyper-scalars under these groups dictates what representations are allowed in the 4D $\mathcal{N}=1$ theory (see the examples in~\cite{McR1}). 

II. Consider now a Calabi-Yau with three moduli $h^{0},h^{1},h^{2}$ such that $\{h^{1},h^{2}\}\rightarrow 0$ induces a divisorial collapse to a genus $g$ curve of $A_{2}$ singularities (i.e. that locally looks like $\mathbb{C}^{2}/\mathbb{Z}_{3}$; see~\cite{KM, KMP} for examples). In the theory on $S^{1}/\mathbb{Z}_{2}$, there are initially three possibilities for assigning parities to the 4D vector fields $A^{1}_{\mu},A^{2}_{\mu}$: Both odd, one even, or both even.  In the first case there are no singularities. In the second case either $h^{1}$ or $h^{2}$ vanishes on the fixed planes and we must embed in an SU(2) YMESGT in a similar fashion as Example I, which involves adding two new functions $h^{3}, h^{4}$, such that the bulk gauge symmetry is $U(1)$ (the other bulk vector is a Maxwell field).  

In the third case, both $\{h^{1},h^{2}\}\rightarrow 0$ on the fixed planes and we must embed in an $SU(3)$ YMESGT, involving four new functions $h^{3},\ldots,h^{8}$, which is broken to $U(1)\times U(1)$ in the bulk. On the fixed planes, this gauge symmetry may be partially enhanced according to two new distinct choices in splitting the new vector multiplets (according to $\mathbb{Z}_{2}$ parity):\\
(i) If all of $A^{3}_{\mu},\ldots,A^{8}_{\mu}$ have odd parity, the boundary symmetry remains broken with six chirals forming the roots of $SU(3)$.\\
(ii) Assigning even parities to two of these vectors yields an enhanced $SU(2)\times U(1)$ symmetry with chiral multiplets in the $\mathbf{2}_{-}\oplus\mathbf{2}_{+}$.         \\
These are the consistent choices one can make.  Again, attempting to have a full $SU(3)$ enhancement on the boundaries removes the additional moduli needed for a non-singular theory.

\section{Yang-Mills-Einstein theories}\label{sec:YMESGT}
What happens when Yang-Mills-Einstein supergravity theories (YMESGTs) are placed on spacetimes of the form $\mathbb{R}^{4}\times S^{1}/\mathbb{Z}_{2}$?  We know generic YMESGTs arise from compactification on generalized Calabi-Yaus, in which an appropriate set of divisors and wrapped membranes have contracted, resulting in an enhanced gauge group $K$ in the bulk spacetime. At the $\mathbb{Z}_{2}$ fixed points in which a subgroup $K_{\alpha}\subset K$ is unbroken due to parity assignments, there will not be any new field content due to the fact that there are no new contractions.\footnote{Of course, for theories with singular Calabi-Yau origin, there must generally be 5D charged hypermultiplets.} However, for any Maxwell vector fields surviving on the fixed planes, the previous MESGT analysis applies.  In Example I of the previous section, an $SU(2)$ gauge symmetry appears at the point in moduli space $h^{1}=h^{2}=h^{3}=0$ with the cycle $C^{1}$ contracted.  If we break this to $U(1)$ on the fixed planes by assigning even parity to $A^{1}_{\mu}$, we require the truncation $h^{1}\rightarrow 0$ there.  Therefore, we are not requiring any new contraction of a cycle at the fixed points, and the ``extra" moduli $h^{2},h^{3}$ are still part of the theory as required. If we try to leave the $SU(2)$ gauge symmetry completely unbroken at the fixed points, then $h^{1},h^{2},h^{3}$ must all be truncated to zero there and we lose the ``extra" moduli that gave us a non-singular theory.  

Finally, consider the $SU(3)$ 5D YMESGT resulting from the Calabi-Yau in Example II of the previous section. The bulk symmetry can be completely broken with eight chiral multiplets remaining on the boundaries; or it can be broken to $SU(2)\times U(1)$ with chiral multiplets in $\mathbf{2}_{+}\oplus\mathbf{2}_{-}$; or finally, broken to $U(1)\times U(1)$ with six charged chiral multiplets (forming the roots of $SU(3)$). 

In general, the 5D gauge group $K$ must be broken on the fixed planes to a proper, rank-preserving subgroup $K_{\alpha}$ (unless all vector fields are projected out).   

 \section{Discussion and Conclusion}  \label{sec:Conclusion}

Vector-coupled supergravity on spacetimes of the form $\mathbb{R}^{4}\times S^{1}/\mathbb{Z}_{2}$ has moduli space singularities at the $\mathbb{Z}_{2}$ fixed points once Maxwell vector fields are allowed to survive there.  This requires new field content such that the theory be embedded in a YMESGT (in a completely Coulombic phase in the bulk spacetime), in which some background scalar functions $h^{I}$ are $x^{5}$-dependent, vanishing on the fixed planes.  On the other hand, a 5D YMESGT with gauge group $K$ requires no new field content with respect to the vectors taking part in the gauging; from the point of view of Calabi-Yau compactifications, the theory is already at singular points.  However, the gauge group must be broken to a proper subgroup of the same rank for consistency.     

Our initial motivation was to understand the framework of 5D supergravity compactified on $S^{1}/\mathbb{Z}_{2}$.  In practice, though concrete examples come from Calabi-Yau compactifications, as the singularity structure for many cases is unambiguous. From the effective supergravity point of view, there is of course an ambiguity in determining the additional field content since a given model represents a class of such Calabi-Yaus, each of which have a distinct singularity structure.  Starting from a particular sugra theory, a change of target space basis maps to an equivalent effective theory, but which may correspond to a different K\"{a}hler cone with a different boundary. From this point of view, we should distinguish sets of bases in sugra according to the resulting singularity structures. 

It should be stressed that in the examples of Calabi-Yau derived theories, we have assumed the validity of the two-stage process of reducing on a CY to five dimensions, then compactifying that theory on $S^{1}/\mathbb{Z}_{2}$, assigning $\mathbb{Z}_{2}$ parities to the 4D vector fields as desired.  We have not offered any seven-dimension singular spaces on which the 11D sugra is to be compactified.  For such a space, $(X_{6}\times S^{1})/\mathbb{Z}_{2}$ with wrapped membranes, one would have to specify how the $\mathbb{Z}_{2}$ action acts.    

Following~\cite{McR2}, we can ask whether the new field content contributes to anomalies, either via inflow or due to quantum loops.  In the cases we have considered in detail, the new field content form real (or zero net charge) representations with respect to the 4D gauge group and therefore do not contribute to quantum anomalies.  More generally, of course, there are situations in which the new hypermultiplets are in representations other than the adjoint (e.g. in the fundamental representation covered in~\cite{NewHiggs, Intriligator:1997pq}) in which case there can be such a contribution.  Next, for both bosonic and supersymmetries, anomaly inflow is governed by the $C_{\alpha\beta\gamma}$ components of the $C_{IJK}$ tensor. These components are precisely those that are generally modified by ``integrating in" new field content.  In Example I of Section~\ref{sec:Examples}, the relevant component, $C_{111}$, is modified so that it vanishes in the case of elementary singularities.  Therefore, the action of the embedding theory is gauge and superymmetrically invariant. Not so for the flop singularity in that example.
\vspace{2mm}\\
\textbf{Acknowledgements}\\
Work supported
 by the European Commission RTN program ``Constituents, Fundamental Forces and Symmetries of the Universe" MRTN-CT-2004-005104 and by INFN, PRIN prot.2005024045-002. \vspace{8mm}\\              
\textbf{\Large{Appendices}}\\
\appendix
\section{Hypermultiplets}\label{sec:Hypermultiplets}
The coupling of hypermultiplets, consisting two spin-1/2 and four scalar fields, to $\mathcal{N}=2$ supergravity was studied in four dimensions in~\cite{BaggerWitten} and in five dimensions in~\cite{hypers}. A common feature is that the hypermultiplet scalars parametrize a quaternionic space such that the total scalar manifold of a hypermultiplet-coupled MESGT is a product $\mathcal{M}=\mathcal{M}_{V}\times \mathcal{M}_{H}$. While there is a well-known collection of homogeneous quaternionic spaces, the non-homogeneous case is not nearly as well understood. The isometries of the total space form the product of isometry groups $Iso(\mathcal{M}_{V})\times Iso(\mathcal{M}_{H})$, and the rigid symmetry group of the Lagrangian is a subgroup $G_{V}\times G_{H}$. To gauge the theory, one must find a group $K$ such that $K\subset G_{V}$ and $K\subset G_{H}$.  In some cases, this requires gauging in $SU(2)_{R}$. The moduli space of the gauged theory is no longer generally a product since the two sectors are gauge-coupled. 

Whereas one can choose not to include hypermultiplets in a strictly 5D supergravity framework, compactification of 11D supergravity on a smooth Calabi-Yau yields $(h^{2,1}+1)$ 5D hypermultiplets, where $h^{2,1}$ is the Hodge number for complex structure moduli of the CY and the additional (universal) multiplet corresponds to the CY volume $\tilde{\mathcal{V}}$~\cite{CYcompactification}.  For singular CYs in which some 2-cycles have contracted, we expect the long-wavelength theory to be described by a YMESGT coupled to charged hypermultiplets as in the previous paragraph.

\end{document}